\documentclass[useAMS,usenatbib]{mn2e}
\usepackage{amsmath}
\usepackage{txfonts}
\usepackage{graphicx}
\usepackage{times}

\begin{document}


\title[Ensemble photo-$z$]{Ensemble photometric redshifts}
\author[Padmanabhan et al]{
  Nikhil Padmanabhan$^1$, 
  Martin White$^{2,3}$, 
  Tzu-Ching Chang$^{4,5}$, 
  J.D. Cohn$^6$, 
  Olivier Dor\'{e}$^{5,7}$, \newauthor
  Gil Holder$^{8}$, \\
    $^{1}$ Departments of Physics and Astronomy, Yale University, New Haven, CT 06511, USA\\
    $^{2}$ Departments of Physics and Astronomy, University of California,
    Berkeley, CA 94720, USA \\
    $^{3}$ Lawrence Berkeley National Laboratory, 1 Cyclotron Road,
    Berkeley, CA 94720, USA\\
    $^{4}$Academia Sinica Institute of Astronomy and Astrophysics, 11F of
    ASMAB, AS/NTU, 1 Roosevelt Rd Sec. 4, Taipei, 10617, Taiwan\\
    $^{5}$ Jet Propulsion Laboratory, California Institute of
    Technology, 4800 Oak Grove Drive, Pasadena,
    CA 91109, USA\\
    $^{6}$ Space Sciences Laboratory and Theoretical Astrophysics Center, University of California,
    Berkeley, CA 94720, USA \\
    $^{7}$ California Institute of Technology, 1200 E California Blvd,
    Pasadena, CA 91125, USA\\
    $^{8}$ University of Illinois, Departments of Physics and Astronomy, Urbana IL 61801 
}
\date{\today}
\pagerange{\pageref{firstpage}--\pageref{lastpage}}

\maketitle

\label{firstpage}

\begin{abstract}
Upcoming imaging surveys, such as LSST, will provide an unprecedented
view of the Universe, but with limited resolution along the line-of-sight.
Common ways to increase resolution in the
third dimension, and reduce misclassifications, include observing a 
wider wavelength range and/or combining the broad-band imaging with 
higher spectral resolution data. The challenge with these approaches
is matching the depth of these ancillary data with the original imaging
survey.
However, while a full 3D map is required for some science, there are many
situations where only the statistical distribution of objects
($dN/dz$) in the
line-of-sight direction is needed.  In such situations, there is no need to
measure the fluxes of individual objects in all of the surveys. Rather a
stacking procedure can be used to perform an ``ensemble photo-$z$''.
We show how  a shallow, higher spectral resolution survey can be used to measure
$dN/dz$ for
stacks of galaxies which coincide in a deeper, lower resolution
survey.  The galaxies in the deeper survey do not even need to appear
individually in the shallow survey.  We give a toy model example to
illustrate tradeoffs and considerations for applying this method.
This approach will allow deep imaging surveys to leverage the high
resolution of spectroscopic and narrow/medium band surveys underway,
even when the latter do not have the same reach to high redshift.
\end{abstract}

\begin{keywords}
  methods:data analysis;
  methods:statistical;
  galaxies:distances and redshifts
\end{keywords}


\section{Introduction}
\label{sec:intro}

The Large Synoptic Survey Telescope (LSST) will be one of the key
astronomical facilities of the next decade.  It will allow us to
map large areas of sky with unprecedented depth in six optical
pass bands ($ugrizY$).
The science enabled by this facility will be revolutionary
\citep[e.g.][]{LSSTbook}.
Fully exploiting these deep sky maps will require information on the
redshifts of the objects.
A redshift estimate can be obtained directly from the
photometry (a ``photo-$z$''), but such redshifts are relatively poor
and can be difficult to obtain for some types of galaxies
\citep[e.g.~see][for recent reviews]{Hil10,Dah13,San14,Rau15}.
Of particular interest here is the use of LSST for studies of
large-scale structure, which heavily impacts cosmology and fundamental
physics.  For such problems the addition of high-quality redshift
information is critical \citep[e.g.][]{New15}.

One can seek to obtain redshifts for individual galaxies or the
redshift distribution for a particularly interesting subsample.
The latter will be the topic of this paper.  Knowledge of $dN/dz$ for a
sample can be used to invert a measured 2D correlation function into a 3D
correlation function \citep{Lim53,Lim54} or to interpret the results of
a cosmic shear experiment \citep{Hoekstra08}.
The most natural method for obtaining $dN/dz$ is to `stack' the photo-$z$s
(or the redshift PDFs) of the galaxies making up the sample.
Another method is to use the fact that objects which are close
on the sky are also likely to be close in redshift.
There is a long history of using such ``cross-correlation'' techniques
to determine $dN/dz$
\citep{SelPee79,Phi85,PhiSha87,Pad07,Ho08,Erb09,Ben10,Ben13,New08,MatNew10,
Sch10,McQWhi13,Mat13,Men13,Sch13,Rah15,Choi16}
and such methods can perform very well.

Unfortunately,  degeneracies in color-type-redshift space are a
 notorious problem with traditional photo-$z$ methods, especially 
when restricted to broad-band optical photometry.  In this case,
galaxies of different types/redshifts can
have the same colors \citep{Benitez00} and thus are indistinguishable. 
These degeneracies are
often easily broken by adding photometry in IR-bands, appropriately 
chosen narrow-band imaging or low-redshift spectroscopy. 
However, it is often challenging to match the
depths of these additional data with the 
original imaging catalog, especially for wide imaging surveys. The key idea
in this paper is the realization that in order to determine redshift 
distributions, it is not necessary to detect individual sources in these 
additional data. For large enough samples, a ``stacked'' measurement can 
constrain the redshift distribution, even if the individual galaxies are
all below the detection threshold. We dub this idea ``ensemble photo-$z$'s''.

This idea is of particular interest since a number of narrow-band imaging/low
resolution spectroscopy large-scale surveys are independently motivated.
For instance, SPHEREx\footnote{http://spherex.caltech.edu}
(Spectro-Photometer for the History of the Universe,
Epoch of Reionization, and Ices Explorer)
is an all-sky spectroscopic survey satellite which will obtain
$R=35-41$ spectra for $0.75<\lambda<3.82\mu$m and $R=110-130$ spectra for
$3.82<\lambda<5.0\mu$m, for a total of 96 bands, for every 6.2 arc second pixel over the entire-sky
\citep{SPHEREx}.  J-PAS\footnote{http://www.j-pas.org/} (Javalambre Physics of the Accelerating
Universe Astrophysical Survey) will cover 8000 deg$^2$ using 56 narrow band filters in
the optical \citep{JPAS}.
PAUS\footnote{http://www.pausurvey.org} (Physics of the Accelerating
Universe Survey) will provide a 100 deg$^2$ 3D maps using 40 narrow band filters covering $4500 < \lambda
< 8500 ~\AA$ on the William Herschel Telescope \citep{PAU2}.
ALHAMBRA \footnote{http://alhambrasurvey.com/} (Advanced Large, Homogeneous Area Medium-Band Redshift
Astronomical Survey) employs 20
contiguous, medium-band filters covering $3500 < \lambda
<9700 \AA$, plus the JHKs near-infrared (NIR) bands, to observe a
total area of 8 deg$^2$ \citep{Alhambra08,Alhambra12,Alhambra14}.

At first glance, none of these surveys are deep enough to provide interesting
additional bands for surveys like LSST. A typical LSST gold-sample galaxy has
an $i$ band magnitude of $\sim 25.3$ and a roughly flat spectrum in the IR. For
SPHEREx, the $5\sigma$ detection limit per frequency element is $\sim19.4$ \citep{SPHEREx};
about 6 magnitudes brighter. However, stacking $\sim 50000$ such LSST galaxies
in SPHEREx (averaging over 5 adjacent SPHEREx bands) should yield a $>10\sigma$ spectrum. Similarly, for J-PAS, taking
the $5\sigma$ detection limit per frequency element to be magnitude $\sim22.5$,
stacking (again averaging over 5 adjacent bands) $\sim 140$ typical LSST gold-sample galaxies would yield a $>10 \sigma$ detection.

As we will see, these stacked spectra encode information about the underlying redshift 
distributions of the objects. Furthermore, since the LSST gold-sample will have
$\sim10^9$ galaxies, it is plausible that, even after dividing into a number of subsamples, 
one would have sufficent numbers of galaxies per subsample to yield stacked detections
in any of these shallower surveys.

The next section lays out the simple idea underlying ensemble photo-$z$'s, 
and then works through two simplified examples. We then conclude with a discussion 
of how one might extend this work, as well as implications for photometric redshift 
calibrations for LSST.

\section{Ensemble photometric redshifts}

We imagine that our data come from two surveys. We assume that the first (denoted
by ${\cal P}$) is a deep, multi-band imaging survey; the prototypical example is LSST, although
one could consider the imaging components of Euclid\footnote{http://sci.esa.int/euclid/} and WFIRST\footnote{http://wfirst.gsfc.nasa.gov/} as well. We imagine this survey
is augmented by a shallower, low-resolution spectroscopic survey ${\cal S}$. Although we assume
spectroscopy below, this could be generalized to a second multi-band imaging survey as well
(with spectroscopy being the infinitesimal band limit). A key element here is that
the majority of objects of interest in ${\cal P}$ are not individually detected in ${\cal S}$.

Start by considering a sample of $N$ galaxies in ${\cal P}$, selected in a small voxel in
observed flux space, and compute the photometric redshifts for these galaxies. Since we have, by
construction, chosen all of these galaxies to have the same observed fluxes, their estimated
photometric redshifts will be the same (ignoring, for now, the scatter due to observational
errors). The accuracy of these redshifts will be intrinsically limited by degeneracies in
flux space -- different templates at different redshifts can produce the same observed fluxes.
This is particularly true for small numbers of filters that span a limited wavelength range.
This problem of interloper redshifts is well known in the photometric redshift literature,
and there have been a number of suggested approaches to reduce or quantify this interloper fraction.
The simplest approach would be to expand the number of filters (with a limit being a
spectrum of the galaxy) and/or the wavelength range to break these degeneracies per object.
A different approach is the idea of clustering redshifts which uses the spatial clustering
of galaxies to constrain the redshift distribution of the ensemble.

Our idea of ensemble photometric redshifts is intermediate between these two approaches : we
will expand the number of ``filters''/wavelength range by augmenting our measurements by
${\cal S}$, but we will not assume that the galaxies are individually detected in ${\cal S}$.
Instead, we use the observation that the average spectrum can be used to constrain
the redshift distribution of the entire sample. Hence ``ensemble photometric
redshifts'' : instead of individually fitting a redshift to each object, we fit the stacked spectrum to measure
the full redshift distribution.

The expected stacked spectrum is just a sum over the $N$ individual
spectra $f(\lambda)$ in ${\cal S}$ centered on the galaxies identified in ${\cal P}$:
\begin{equation}
  f_{\rm av}(\lambda) = \sum_{i=1}^{N} f_i(\lambda) \,\,,
\end{equation}
where $\lambda$ is in the observer frame. If we imagine that galaxy spectra are well described by a
relatively small number of templates $F_{\alpha}(\lambda,z)$, we can rewrite the above as
\begin{equation}
  f_{\rm av}(\lambda) = \sum_{i=1}^{N} A_i F_{\alpha_i} (\lambda, z)
\end{equation}
where the normalization $A_i$ depends on the luminosity of the object. We can simplify this
further using the fact that we selected galaxies from a narrow voxel in {\it observed} flux
space. This implies that all objects with the same spectrum $F_{\alpha}$ at the same redshift
must have the same normalization, and we are free to absorb this normalization into the definition
of the galaxy template $F_{\alpha}(\lambda, z)$. The stacked spectrum can now be written as
\begin{equation}
  f_{\rm av}(\lambda) = \sum_{\alpha} \int\,dz\, \left(\frac{dN}{dz}\right)_{\alpha} F_{\alpha}(\lambda, z)\,
  \label{eq:fav}
\end{equation}
where $(dN/dz)_\alpha$ is the redshift distribution of galaxies of type $\alpha$. The ensemble
photometric redshift problem is now analogous to regular photometric redshifts - we consider
maximizing the likelihood $L(f_{\rm av} | \{ (dN/dz)_{\alpha} \})$ (or determining the corresponding
posterior distribution).

Although the above expression does not explicitly list the dependence on the data from the original
photometric survey ${\cal P}$, this is implicit in our choice of
template spectra and their normalizations. This is possible because we started by selecting galaxies
in a narrow voxel in flux-space from ${\cal P}$. While this is a useful simplification,
it is possible to extend this to more complicated selections by augmenting the likelihood to
$L({\cal D}_{\cal P}, f_{\rm av} | \{z_i, \alpha_i, (dN/dz)_{\alpha} \})$, where ${\cal D}_{\cal P}$
represents the data from ${\cal P}$ and we now estimate the individual $z_i$ and $\alpha_i$ jointly
with the redshift distributions.

While we chose to work with a stacked spectrum above, one could imagine directly fitting the
observed fluxes of the individual galaxies in ${\cal S}$; this would be the optimal approach if
one had an accurate model of the errors in the fluxes. On the other hand, working with the stacked
fluxes allows one to e.g. diagnose template mismatches or an inadequate error model. Again, the choice
of using a stacked spectrum is not essential to this discussion but is a useful mental model of
the idea of an ensemble photo-$z$; the key idea is to jointly fit all observations to constrain the
redshift distributions.

\subsection{Example 1 : Breaking degeneracies}

\begin{figure}
  \centering
  \includegraphics[width=0.9\columnwidth]{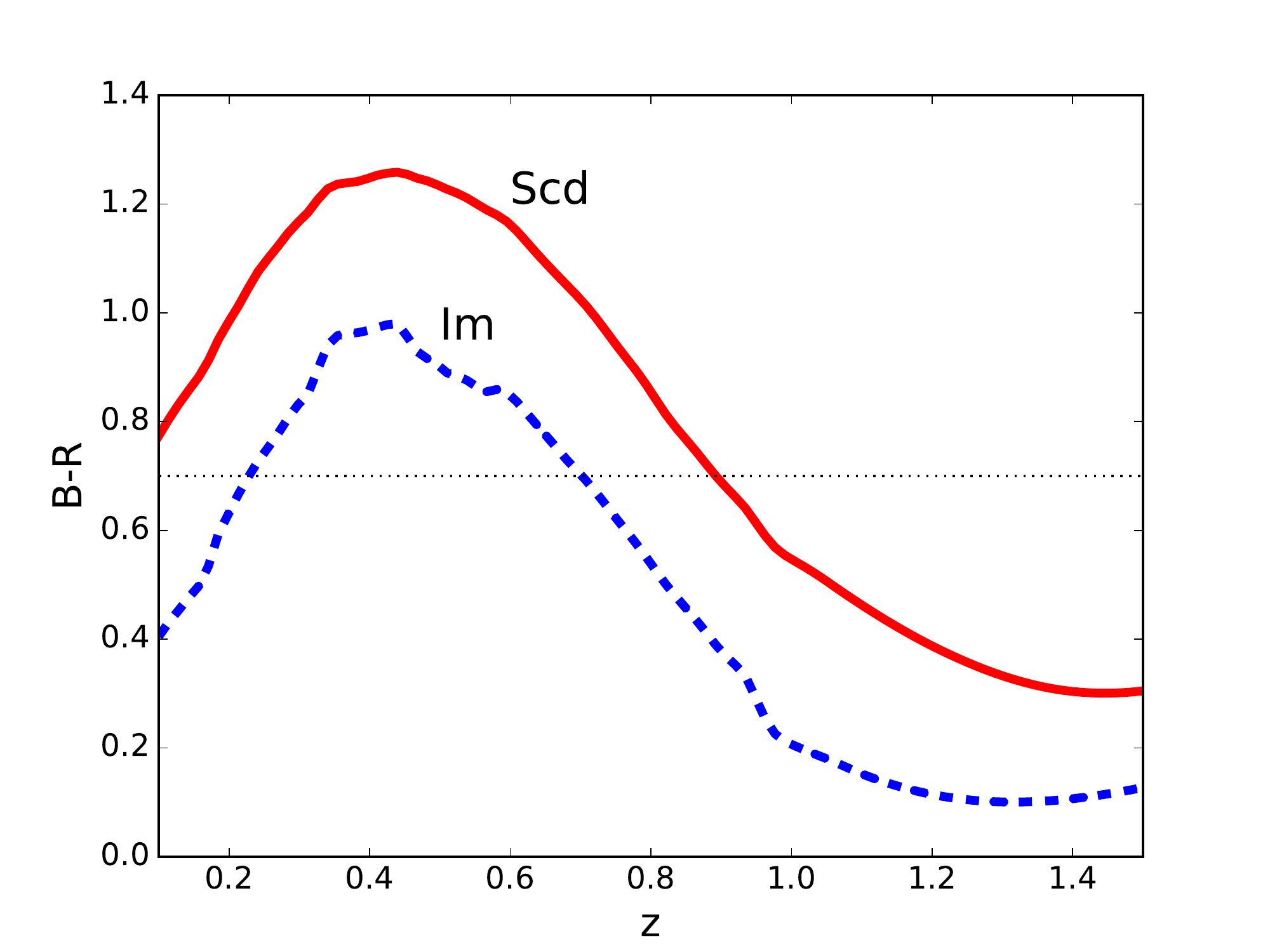}
  \caption{The $B-R$ color of a spiral (Scd) and irregular (Im) galaxy, highlighting
  the color-redshift degeneracies. The galaxy templates are from \citet{Benitez04}
  and are normalized to have the same observed flux.
  We approximate the $B$ and $R$ filters as tophat filters centered at $4450~\AA$
  and $6580~\AA$ with widths of $700~\AA$ and $1000~\AA$ respectively. An example
  color-redshift degeneracy is highlighted by the dotted line at $B-R=0.7$; this
  color is consistent with the irregular galaxy at $z\sim0.2$ and $\sim0.7$ and
  the spiral galaxy at $z\sim0.9$.}
  \label{fig:twocolor}
\end{figure}

We start with a simple example that demonstrates how a stacked spectrum can break degeneracies
in redshift distributions. We imagine a sample of galaxies in ${\cal P}$ that are drawn from two populations :
spiral and irregular galaxies (we use the \texttt{Scd\_B2004a} and
\texttt{Im\_B2004a} templates in \citet{Benitez04}). These galaxies
are observed through two simple top-hat filters at $4450~\AA$ and
$6580~\AA$ (similar to $B$ and $R$ filters). Fig.~\ref{fig:twocolor}
shows the $B-R$ color tracks for these galaxies as a function of
redshift. For galaxies with an observed $B-R$ color of 0.7 (shown by
the dotted line) and a unit (in arbitrary units) $B$-band flux, we
observe a three-fold redshift/type degeneracy - the irregular galaxy
could be at redshifts $\sim0.2$ and $\sim0.7$, while the spiral galaxy
could be at $\sim0.9$. Photometric errors would naturally broaden
these distributions. 


\begin{figure}
  \centering
  \includegraphics[width=0.9\columnwidth]{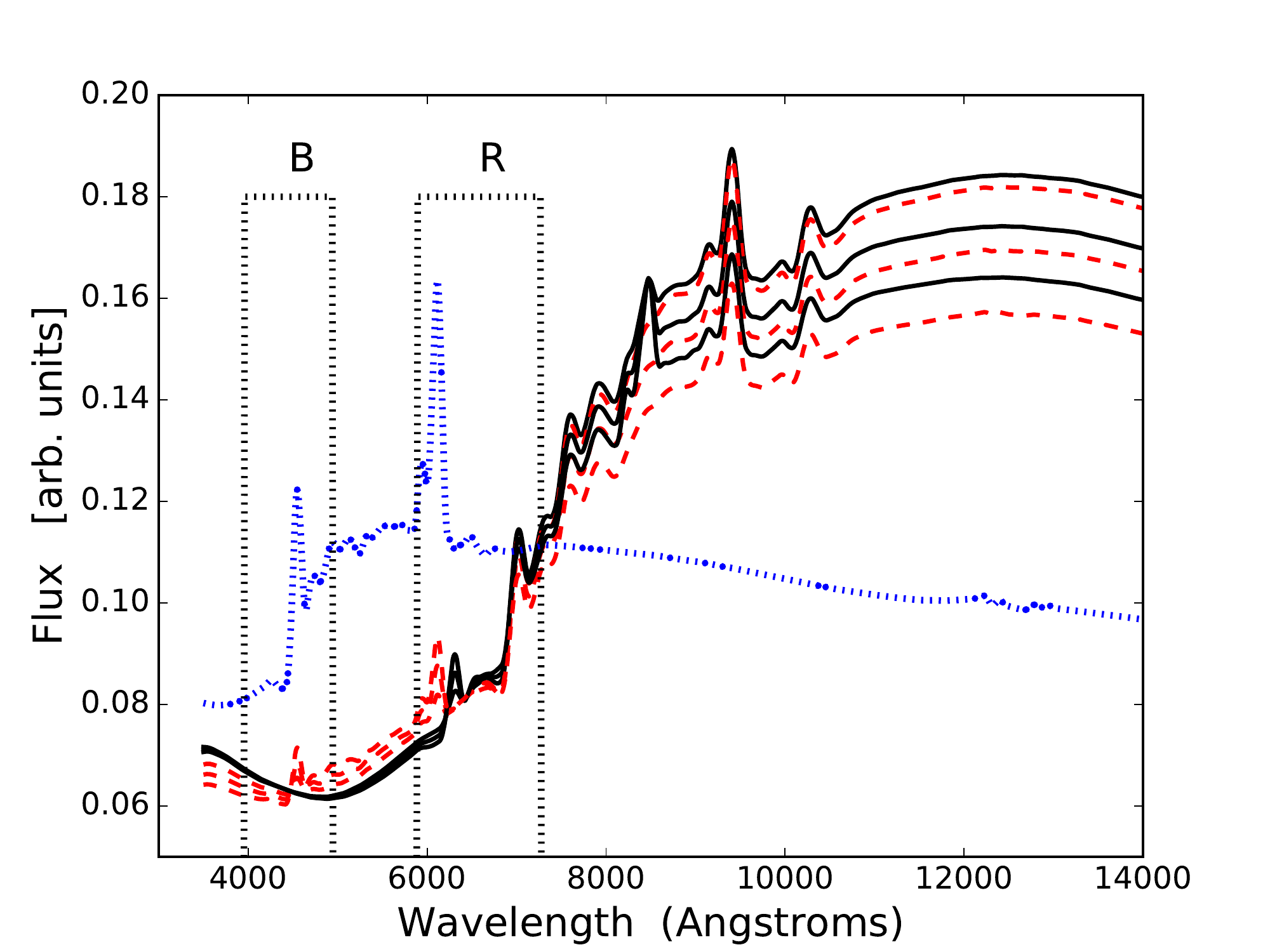}
  \caption{A demonstration that a stacked spectrum can be used to
  break degeneracies in photometric redshifts. From top to bottom,
  the lines show the expected stacked spectrum of
  spiral (Scd) galaxies with 10\%, 20\% and 30\% contamination from
  a population of irregular (Im) galaxies. In all cases, the spiral
  galaxies are at a redshift of $\sim$0.9; the solid (black) lines have
  the irregular galaxies at $z\sim0.7$, while the dashed (red) lines
  have the irregular galaxies at $z\sim0.2$. By construction, all of these
  galaxies have the same $B-R$ color, and the same $R$-band magnitude.
  Also plotted for reference [dotted] is the irregular galaxy spectrum at $z=0.2$
  (with an arbitrary vertical shift for clarity), and the nominal $B$
  and $R$ bands we use.}
  \label{fig:spectrum1}
\end{figure}

We now imagine observing this population of galaxies with a low-resolution spectroscopic survey.
We follow our example above, selecting galaxies with a fixed $B-R = 0.7$ color and
unit $R$-band flux. Fig.~\ref{fig:spectrum1} plots the predicted spectrum for different
admixtures of types and redshifts. For definiteness, we assume that the spiral
galaxies form the dominant population, with the irregulars being contaminants.
For this particular example, we find that the $\lambda > 1 \mu m$ part of the
spectrum can determine the overall contamination fraction. The differences
between contaminants at different redshifts (here $z\sim0.2$ and $0.7$) are
smaller, although with enough S/N, one can clearly start to distinguish 
these cases. The details are clearly specific to the case we have
chosen here, but this demonstrates that an averaged spectrum can 
break degeneracies in photometric redshift distributions.

Furthermore, this figure allows us to schematically
understand the depth requirements for the spectroscopic survey - given two degenerate (in $\cal P$)
populations of galaxies, one needs to be able to distinguish between the stacked spectra in ${\cal S}$.
As a numerical example, we suppose that stacking $10^4$ LSST galaxies in
${\cal S}$ yields a 10\% detection of flux per frequency element and that ${\cal S}$
has $\sim 25 - 100$ frequency elements. Then we should be able to
detect 1-2\% differences in (total) flux in the spectrum, and distinguish
between the different spectra in Fig.~\ref{fig:spectrum1}.
The LSST gold sample ($i < 25$) is expected to contain
$\sim 4 \times 10^9$ galaxies, which suggests that one could conceptually
break the sample into $\sim 10^5$ voxels in magnitude space, each with 
sufficient galaxies to stack in the spectroscopic survey. 
The above estimates are just meant to be illustrative and to demonstrate that 
such an approach is feasible in principle.

\subsection{Example 2 : Fitting redshift distributions}

\begin{figure}
  \centering
  \includegraphics[width=0.9\columnwidth]{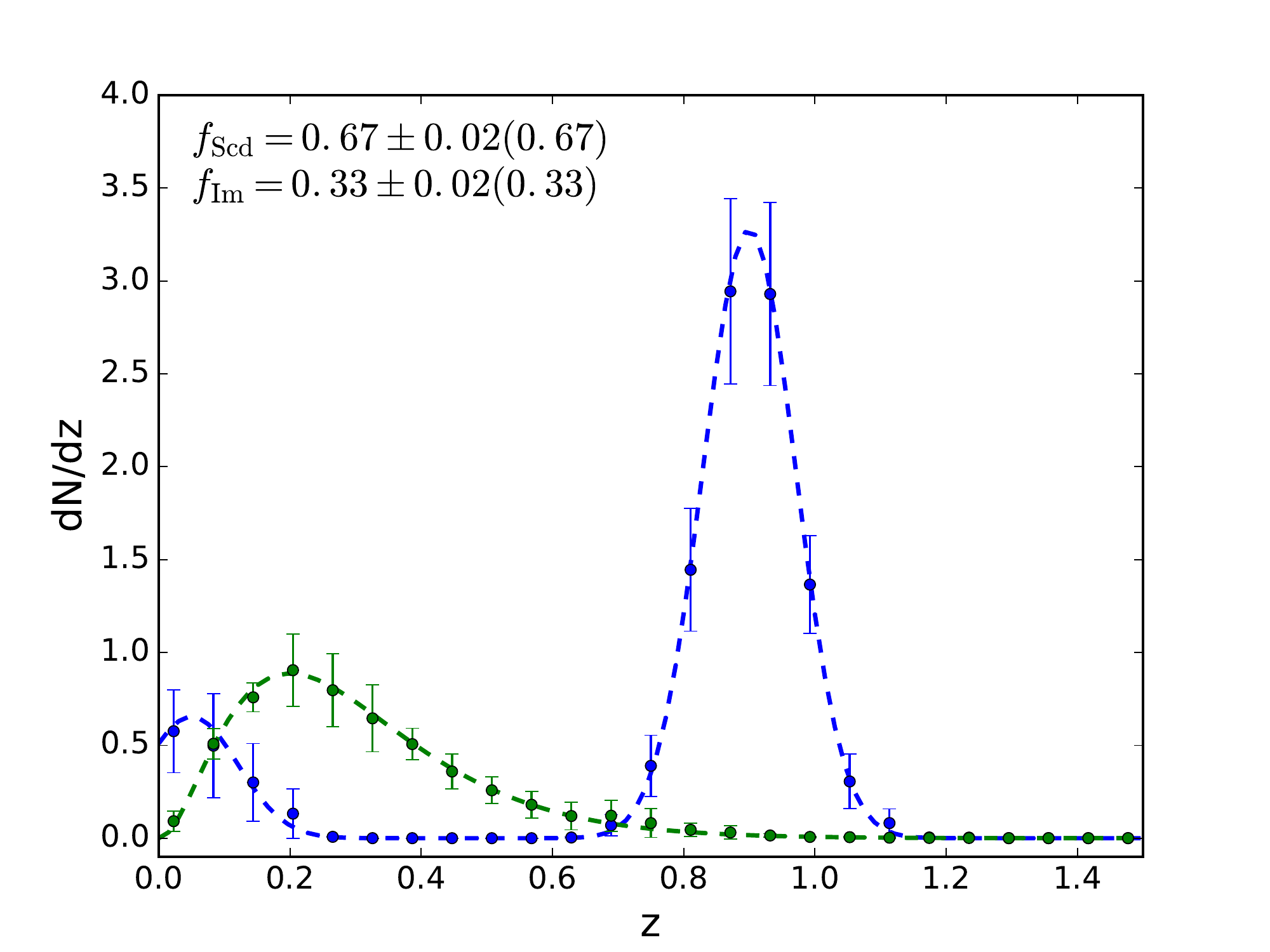}
  \caption{A test showing the recovered redshift distribution using
    stacked spectra. The dashed lines are the input redshift distributions
    for spiral (blue) and irregular (green) redshifts. The points show
    the recovered redshift distributions (averaged over 50 simulations),
    while the errorbars show the uncertainty expected for a single
    realization. Also shown are the fractions of spiral and irregular
    galaxies with the input values in parentheses. See the text for more
  details on the exact simulations. }
  \label{fig:dndz}
\end{figure}

The previous section demonstrated how the stacked spectrum could
break degeneracies in photometric redshifts. We extend this idea
to measuring the redshift distribution here. 
As in the previous
section, we consider a mixture of spiral and irregular galaxies using
the same templates used previously.
The assumed redshift distributions are shown in Fig.~\ref{fig:dndz}.
The forms we chose reflect the color-redshift degeneracies
seen in Fig.~\ref{fig:twocolor}; in particular, note the ``contamination''
of low-redshift spirals and irregulars.
We imagine these galaxies are selected to have the same $R$-band
flux, they have perfectly measured $B-R$ colors, and that this selection
yields $10^6$ galaxies.

We now combine the above redshift distributions (for spirals and irregulars)
into a single $B-R$ color distribution.
We then split the sample into $\Delta (B-R)=0.1$
color bins from $B-R=0$ to $1.3$. For each of these color bins, we compute the
stacked spectrum of all the galaxies in the bin. Fig.~\ref{fig:twocolor} shows
that, in general, these stacked spectra are the combination of spiral
and irregular galaxies at two different redshifts.
We assume that stacking $10^4$ galaxies yields a 10\% measurement of flux per 
frequency element in the spectroscopic survey, and we scale this error 
by $\sqrt{10^4/N}$ where $N$ is the number of galaxies in the $B-R$ bin 
under consideration. We also assume that the stacked spectrum is measured
from $5000\AA$ to $14000\AA$ with $R\sim100$, which yields $\sim70$ frequency 
elements. The inputs to our algorithm are the $B-R$ color distribution and the
stacked spectrum in each $\Delta(B-R)$ color bin. As with template-based photo-$z$
codes, we assume we have a complete set of spectral templates (in this case,
the templates for spiral and irregular galaxies).

We parametrize the redshift distributions of each individual
population (spiral or irregular) by step-wise constant distributions in
$z$ with 100 bins from $z=0$ to $z=1.5$. Given a bin in $B-R$, Fig.~\ref{fig:twocolor}
shows that only a small fraction of these redshift bins will have a consistent $B-R$ color.
These redshift bins are the input variables to a least-squares
fit to the observed (stacked) spectrum using Eq.~\ref{eq:fav}.
We impose additional constraints that $(dN/dz)_{\alpha, b} \ge 0$ and
that $\sum_z (dN/dz)_{\alpha, b} = 1$, where $b$ indexes the color bin,
and the latter constraint normalizes the redshift distribution.
After computing these redshift distributions over the individual
color bins, we combine these using
\begin{equation}
  \left( \frac{dN}{dz} \right)_{\alpha} = \sum_{b} N_b \left( \frac{dN}{dz} \right)_{\alpha, b} \,,
\end{equation}
where $N_b$ is the number of galaxies in the $b^{\rm th}$ color bin.

Fig.~\ref{fig:dndz} shows the redshift distribution recovered using the above
procedure, averaging over the results of 50 simulations. Although we estimate
the redshift distribution over 100 bins, the values between adjacent bins
are highly covariant (since the spectra do not have the S/N to distinguish
between small changes in redshift). We therefore average neighbouring bins
to produce the figure shown. We also compress our results to the fractions
of spiral and irregular galaxies, to mimic the case where the shape of the
redshift distributions might be known (or well-constrained).
We see that this simple procedure recovers the correct fractions
of spiral and irregular galaxies as well as their redshift distributions. While
this is a toy example, it illuminates the utility of these stacked spectra.

\section{Discussion}
\label{sec:discussion}

We introduce the idea of ``ensemble'' photometric redshifts as a tool 
to constrain photometric redshift distributions. The idea is a simple extension
of current template-based photometric redshift codes, and uses the fact that 
the shape of a stacked spectrum encodes information about the redshift 
distribution of the galaxies being stacked. The advantage of this approach 
is that the individual galaxies no longer need to be detected in the 
second survey, opening up the possibilities of using planned low-resolution shallower
spectroscopy surveys like J-PAS, PAU and SPHEREx to calibrate deeper surveys
like DES and LSST. An important point is that the next generation of 
imaging surveys will have samples of $\sim 10^9$ galaxies, which allows 
one to build large numbers of subsamples, each of which have sufficient 
numbers of galaxies to stack in the shallower survey.
We outline the idea in this paper, and discuss some 
simple examples demonstrating how these stacked spectra can be used to 
break photometric redshift degeneracies and measure redshift distributions. 
These examples are meant to be illustrative; future work will be needed
to understand the signal to noise for realistic galaxy distributions for future
surveys. 

An aspect of the ensemble photo-$z$ method is that one simultaneously fits both the 
individual photometric redshifts and the redshift distribution. We outline 
a simplified algorithm here, where we imagine splitting the original sample
into voxels in magnitude space. This problem has also recently been considered by 
\citet{Leistedt16} who discuss a more general approach to this problem; their 
algorithm can be easily extended to include constraints from the stacked spectrum. 
We expect that future work will also consider optimal algorithms for the 
next generations of surveys.

Our approach here has been to stack sources to get a detection in the shallower
survey. Clearly, if one has well characterized errors, it is clear that the 
same information can be recovered by fitting observed fluxes (even if they 
are individual non-detections). We were however motivated by the fact that 
stacking the galaxies allows us to develop better intuition for the process. It 
also opens up possibilities for detecting mismatches in photometric redshift 
templates used, which could be folded back into photometric redshift codes.
It should be emphasized that this entire process requires that one can average
down the noise in the shallow spectroscopic survey, which will impose requirements 
on the data reduction and calibration. 

Although the examples presented here used simple models for galaxy populations,
our formalism is straightforward to extend to more complex cases. One such complication
is the effect of dust, which will smear a single population at a fixed redshift
along a line of extinction. We can imagine introducing parameters describing the
scatter in extinction/reddening as well as its direction into our model, and
then simultaneously fitting/marginalizing these with the redshift distributions.
We defer a detailed study of this and other real-world complications to future work.

The problem of determining the photometric redshifts for the next generation
of surveys is still an open question. It has been long recognized that increasing 
the wavelength coverage can improve photometric redshifts; however, it is 
normally assumed that these additional data should be matched in depth to the 
primary survey. We point out that, for the specific problem of determining 
the redshift distribution, this is not necessary, opening up the possibility 
for alternative/easier routes to calibrating photometric redshift distributions.

\vspace{0.2in}
NP is supported in part by DOE DE-SC0008080. T.-C. C. acknowledges
support from MoST grant 103-2112-M-
001-002-MY3 and the Simons Foundation.
This work was begun and completed at the Aspen Center for Physics, which is supported
by National Science Foundation grant PHY-1066293.
This work made extensive use of the NASA Astrophysics Data System and
of the {\tt astro-ph} preprint archive at {\tt arXiv.org}. Part of the
research described in this paper was carried out at the Jet Propulsion
Laboratory, California Institute of Technology, under a contract with
the National Aeronautics and Space Administration.


\newcommand{\aj}{AJ}
\newcommand{\apj}{ApJ}
\newcommand{\apjs}{ApJ Suppl.}
\newcommand{\mnras}{MNRAS}
\newcommand{\araa}{ARA{\&}A}
\newcommand{\aap}{A{\&}A}
\newcommand{\pre}{PRE}
\newcommand{\prd}{Phys. Rev. D}
\newcommand{\apjl}{ApJL}
\newcommand{\physrep}{Physics Reports}
\newcommand{\nat}{Nature}

\bibliographystyle{mn2e}
\bibliography{ms}

\label{lastpage}
\end{document}